\documentclass[11pt]{article}

\usepackage{amsfonts,latexsym,eucal,color,graphicx,epsfig,amssymb,amsmath}

\newcommand{\be}{\begin{eqnarray}}
\newcommand{\ee}{\end{eqnarray}}
\newcommand{\pd}{\partial}

\newcommand{\nn}{\nonumber}

\newcommand{\lan}{\langle}
\newcommand{\ran}{\rangle}
\newcommand{\dalm}{\kern1pt\vbox{\hrule height 0.9pt\hbox{\vrule width 0.9pt\hskip 2.5pt\vbox{\vskip 5.5pt}\hskip 3pt\vrule width 0.3pt}\hrule height 0.3pt}\kern1pt}
\newcommand{\ie}{{\it i.e.}}
\newcommand{\eg}{{\it e.g.}}

\newcommand{\et}{{\it et al.}}

\newcommand{\vs}{{\it v.s.}}
\newcommand{\dd}{{\rm d}}

\makeatletter

\@addtoreset{equation}{section}
\makeatletter

\begin{document}

\begin{titlepage}

\begin{flushright}
{\small {\tt arXiv:1012.2374 [gr-qc]}} \\
\end{flushright}
\vspace{.8cm}
\begin{center}
{\Large  {\bf Particle Creation by Naked Singularities in Higher Dimensions}}\\
\vspace{1cm}
Umpei Miyamoto${}^\ast$, \; Hiroya Nemoto, \; and \; Masahiro Shimano\\
\vspace{.3cm}
{\it {\small Department of Physics, Rikkyo University, Tokyo 171-8501, Japan}}\\
\vspace{.3cm}
{\tt {\small ${}^\ast$umpei@rikkyo.ac.jp}}
\end{center}
\vspace{.2cm}
\begin{abstract}
Recently, the possibility was pointed out by one of the present authors and his collaborators that  an effective naked singularity referred to as ``a visible border of spacetime'' is generated by high-energy particle collision in the context of large extra dimensions or TeV-scale gravity. In this paper, we investigate the particle creation by a naked singularity in general dimensions, while adopting a model in which a marginally naked singularity forms in the collapse of a homothetic  lightlike pressureless fluid. We find that the spectrum deviates from that of Hawking radiation due to scattering near the singularity but can be recast in quasi-thermal form. The temperature is always higher than that of Hawking radiation of a same-mass black hole, and can be arbitrarily high depending on a parameter in the model. This implies that, in principle, the naked singularity may be distinguished from a black hole in collider experiments.
\end{abstract}

\vspace{0.6cm}

\tableofcontents

\end{titlepage}

\section{Introduction}
\label{sec:intro}

The higher-dimensional scenarios with large~\cite{ArkaniHamed:1998rs} and warped~\cite{Randall:1999ee} extra dimensions were proposed
to resolve or reformulate the hierarchy between the gravitational and electroweak interactions. In these scenarios the $d$-dimensional ($d>4$) Planck mass $M_P$ is related to that in 4-dimension $M_{P(4)} (\sim 10^{19} \; {\rm GeV})$ by $ M_{P(4)}^2 \sim L^{d-4} M_P^{d-2} $, where $L$ is the size of an extra dimension. Thus, if $L$ is large enough, the fundamental Planck mass $M_P$ can be as low as a few TeV. For instance, such a small Planck mass is realized if $L \sim 0.1$ cm when $d=6$, $L \sim 10^{-7}$ cm when $d=7$. If the Standard Model particles but the gravitons (and possibly other unobserved particles) are confined to our 3-brane these scenarios are consistent with all current observations.

One of the most striking predictions of such scenarios is the productions of a large number of mini black holes in high-energy particle collisions~\cite{Argyres:1998qn}. A simplified picture of the black hole production is put in the following way. The size of a black hole is characterized by Schwarzschild radius $r_H$, scaling with its mass as $r_H \propto M_{BH}^{1/(d-3)}$. If colliding particles (partons) have a center-of-mass energy $ M_{BH}$ above a threshold energy of order $M_P$ and an impact parameter less than the Schwarzschild radius, a black hole of mass $M_{BH}$ is produced. In other words, the total cross section of black hole production is given by $\sigma_{BH} \simeq \pi r_H^2$. The black holes so produced will decay thermally via the Hawking radiation~\cite{Hawking:1974sw} and be detected in terrestrial collider experiments such as CERN Large Hadron Collider and in ultrahigh energy cosmic rays. Such possibilities have been extensively studied and known to give rise to rich phenomenology (\eg, see \cite{Kanti:2008eq} for a recent review).

Quite recently, one of the present authors and his collaborators pointed out another interesting possibility~\cite{Nakao:2010az} of the high-energy particle collisions in the TeV-scale gravity. They argued that effective naked singularities called {\it the visible borders of spacetime} would be generated by high-energy particle collisions. A border of spacetime, proposed in \cite{Harada:2004mv} originally, is defined as a domain of spacetime where the curvature becomes trans-Planckian and acts as an border (or boundary) of classical sub-Planckian regimes. A simplified picture of the generation of a visible border is as follow. Suppose colliding particles have a center-of-mass energy above $M_P$, just like the black hole production mentioned above. Then, suppose that the impact parameter is small enough to make the energy density of the colliding region become trans-Planckian but the impact parameter is {\it larger} than the Schwarzschild radius. In such a case, if any, the curvature around the colliding region becomes trans-Planckian through the Einstein equation, but the horizon will not form\footnote{It was argued in~\cite{Giddings:2004xy}, replying to the claim raised in~\cite{Rychkov:2004sf}, that the curvature singularity at the collision event described by the Aichelburg-Sexl waves can be avoided if one takes into account the finite extension of colliding particles as a quantum effect. On the other hand, the argument of visible-border production in~\cite{Nakao:2010az} does not depend on how to describe the collision but on a dimensional estimation of the energy density and curvature around the collision event. Furthermore, the argument takes into account the possible quantum uncertainty of particles' position. Thus, at this point, we believe that the visible-border production would not be avoided even if one takes into account the semiclassical effects such as the uncertainty. Of course, however, it would be important to verify further the reality of the visible-border production from various semiclassical points of view. This paper, in which another semiclassical effect --- the particle creation --- is investigated, can be regarded as a part of such verifications. The authors thank an anonymous referee for making us think of this point.}. Thus, the trans-Planckian domain of spacetime, \ie, the border of spacetime, is visible or naked to outer observers. Paper~\cite{Nakao:2010az} showed with a simple argument that such phenomena can occur in collider experiments, that is regarded as an effective violation of Cosmic Censorship Hypothesis~\cite{Penrose:1969pc}. In such a visible border production, in contrast to the black hole production, the trans-Planckian regime is exposed to observers, and therefore an arena of quantum gravity could be provided.

It would be fare to say that the production of visible borders (or naked singularities) as well as the black holes in collider experiments is just a possibility: we do not have any rigorous quantum theory of gravity to predict phenomena near the Planckian regime. Nevertheless, it would be important to predict possible phenomena with using classical/semiclassical tools available. The study of semiclassical effects during naked singularity formation (in 4 dimensions) has a relatively long history, which goes back to the seminal works by Ford and Parker~\cite{Ford:1978ip} and by Hiscock \et~\cite{Hiscock:1982pa}. (An incomplete list of studies in this direction is~\cite{Barve:1998ad,Harada:2000me}. See \cite{Harada:2001nj} for a review.) Typically, when a (strong) global naked singularity forms, which violates even the weak version of the Cosmic Censorship Hypothesis~\cite{Penrose:1969pc}, the power of particle creation diverges at the Cauchy horizon (if estimated without backreactions).

Due to the universality of particle creation by naked singularities, one can expect that similar phenomena will occur in higher dimensions, which is addressed in this paper. However, we should note several points. Firstly, the generation of visible borders or naked singularities in large extra dimension scenarios would be a highly asymmetric phenomenon: the gravity propagates in all dimensions; the Standard Model particles are confined to the 3-brane; the non-zero impact parameter is essential. It seems difficult to model the visible-border formation by a known solution to Einstein equations. Therefore, in this paper, as a first step, we adopt the Vaidya solution describing the spherically symmetric naked singularity formation due to the accretion of a null dust fluid. Secondly, the spectra of particle creation will provide important informations in order to identity what the products of collision are in experiments/observations. However, there is a fundamental problem in estimating the spectrum of created particles (\ie, the Bogoliubov coefficients) when the singularity is globally naked: we do not know how to impose boundary conditions on a quantum field at the singularity. Thus, to avoid such an ambiguity we adopt the model first considered in \cite{Hiscock:1982pa}. This model describes the formation of a {\it marginally} naked singularity, in which the singularity is observable only within the event horizon (so, it violates only the strong version of the Cosmic Censorship Hypothesis~\cite{Penrose:1969pc}). In this spacetime the Cauchy horizon and event horizon coincide, and therefore one has to impose no boundary conditions at the singularity.

This paper is organized as follow. In Sec.~\ref{sec:background}, we introduce the homothetic or self-similar Vaidya solution as the background spacetime describing the formation of the marginally naked singularity. In Sec.~\ref{sec:KG}, we consider the creation of massless scalar particles, and find that the spectrum can be recast in quasi-thermal form. In Sec.~\ref{sec:temperature}, we investigate the physical consequences of the result obtained in the preceding section, such as the implications of a modified temperature. The final section is devoted to conclusion. We use the Planck units unless otherwise noted, in which $c=G_d=\hbar=k_B=1$. 

\section{Background spacetime}
\label{sec:background}

\subsection{Homothetic Vaidya collapse}
\label{sec:vaidya}

We consider a $d$-dimensional ($d\geq 4$) spherically symmetric spacetime of which line element is given by
\be
&&
	\dd s^2
	=
	g_{ab} \dd x^a \dd x^b
	=
	-
	\left(
		1-\frac{2 m(v)}{r^{d-3}}
	\right) \dd v^2
	+
	2 \dd v \dd r
	+
	r^2 \dd\Omega_{d-2}^2\ ,
\label{eq:metric}
\ee
where $\dd \Omega_{d-2}^2$ is the line element of a unit ($d-2$)-sphere. For an arbitrary $m(v)$, the following stress-energy tensor solves the Einstein equation,
\be
	T_{ab} = \frac{d-2}{8\pi} \frac{\dd m}{ \dd v } k_a k_b\ ,
\ee
where $k_a := -\delta_a^v $ is an ingoing null vector. Here, we choose the mass function $m(v)$ as
\be
	m(v)
	=
	\left\{
		\begin{array}{lll}
		0,	&	v < 0	&	\mbox{(I)}\\
		\mu v^{d-3},	&	0 \leq v < v_0	&	\mbox{(II)}\\
		M,	&	v > v_0 &	\mbox{(III)}
		\end{array}
	\right. \ ,
\label{eq:m(v)}
\ee
where $\mu$, $v_0$, and $M$ are all positive constants. Then, the metric (\ref{eq:metric}) describes the gravitational collapse of a null dust fluid or pressureless radiation flowing toward the center from the asymptotic region. The above choice of $m(v)$ results in the homothety or self-similarity of the solution. It is known that a central singularity appears at $(v,r)=(0,0)$, and whether the singularity is naked or not depends on the parameters. Usually, the mass parameter $M$ of the Schwarzschild region III is chosen as $M=\mu v_0^{d-3}$, which results in that the local event horizon in the Vaidya region II\footnote{We mean by the local event horizon the event horizon when the spacetime is entirely filled with the null dust.} is connected to that in region III. For now, we do not impose such a condition, and three parameters ($\mu, v_0, M$) are left to be independent. We impose a different condition to realize a marginally naked singularity in the next subsection. It is noted that the physical mass defined in the asymptotic region is given by~\cite{Myers:1986un},
\be
	M_{BH}
	=
	\frac{ (d-2) \Omega_{d-2} }{ 8\pi } M\ ,
\label{eq:MBH}
\ee
where $\Omega_{d-2} = 2\pi^{(d-1)/2}/\Gamma[(d-1)/2]$ is the volume of the unit $(d-2)$-sphere.

\begin{figure}[t!]
	\begin{center}
		\setlength{\tabcolsep}{ 30pt }
		\begin{tabular}{ ccc }
			\includegraphics[width=4.5cm]{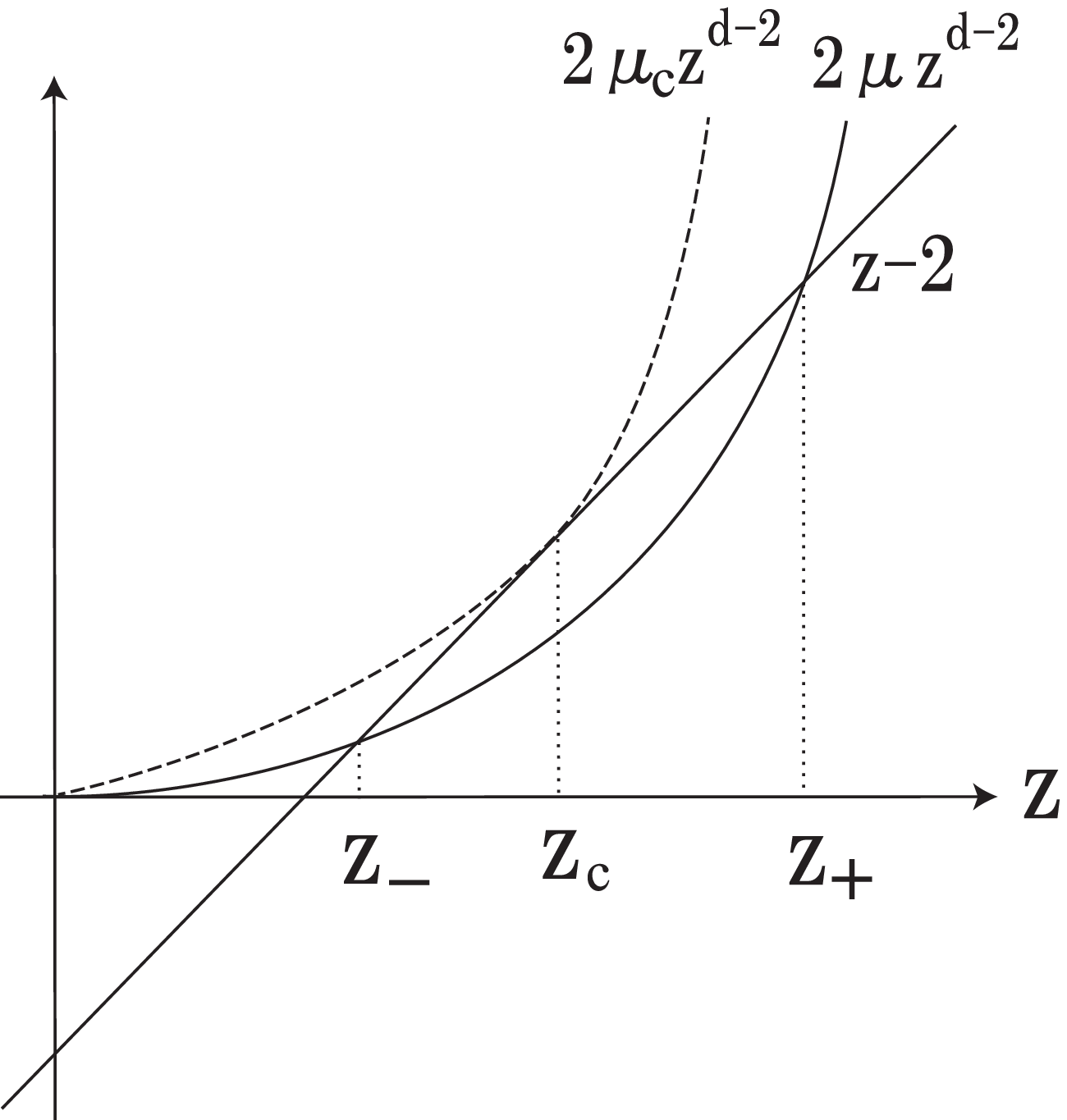} &
			\includegraphics[width=6cm]{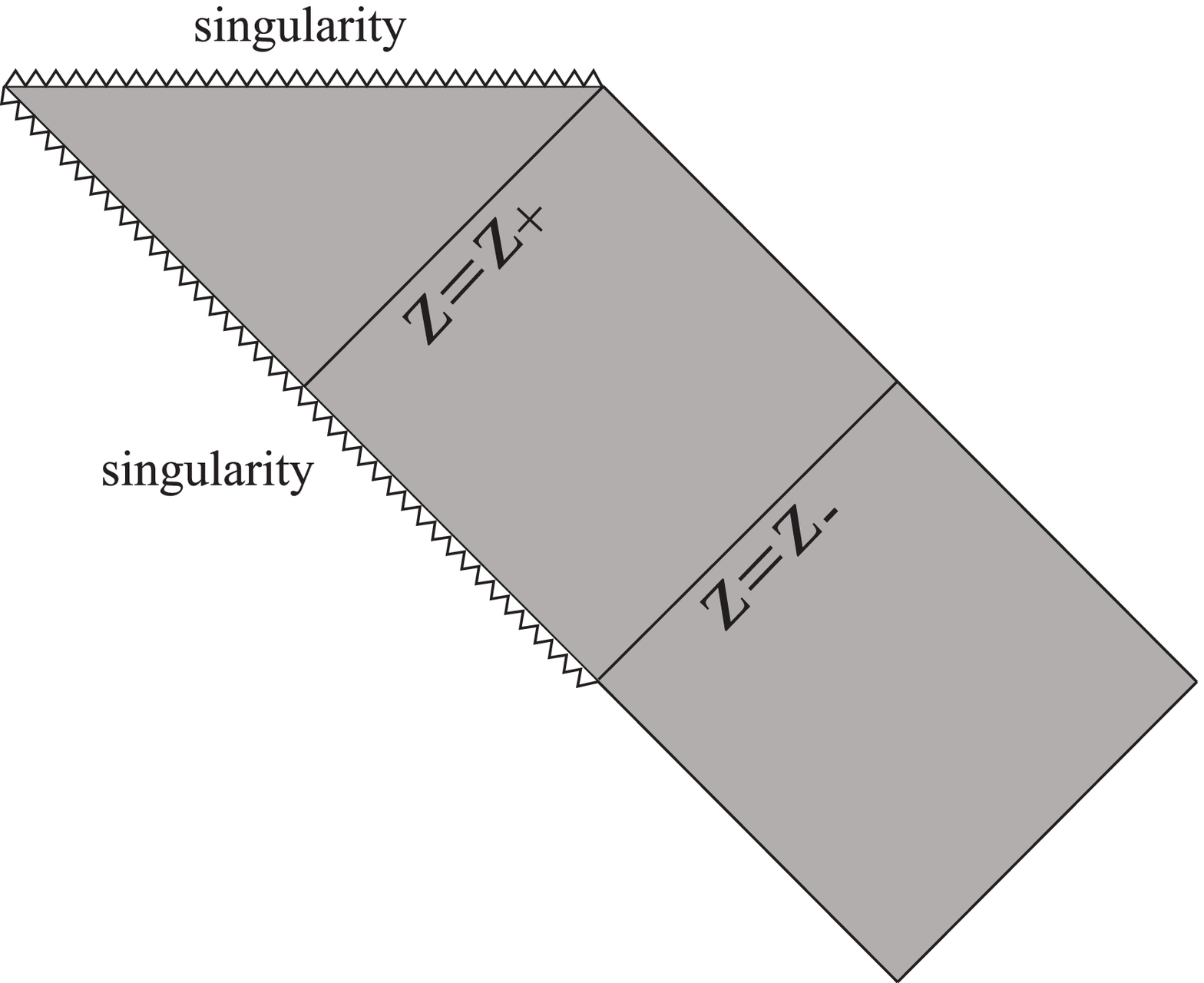} \\
			(a) & (b) \\
		\end{tabular}
	\caption{\textsf{(a) Two roots of $h(z)= 2\mu z^{d-2}-(z-2) =0$ determines the Cauchy horizon $z=z_-$ and event horizon $z=z_+$. In the limit of $\mu \to \mu_c$ the two roots degenerate, \ie, $z_{\pm} = z_c$ . (b) A conformal diagram of the spacetime entirely filled with the null dust fluid.}}
	\label{fg:hz}
	\end{center}
\end{figure}

In region II we introduce a new time coordinate $\zeta$ and similarity coordinate $z$ as
\be
	\zeta
	:=
	\ln \left( \frac{v}{v_1} \right)\ ,
\;\;\;\;\;\;
	z
	:=
	\frac{v}{r}\ ,
\ee
where $v_1$ is a positive constant.\footnote{This parameter $v_1$ is introduced just for the dimensions of line element to be correct, but any results in this paper do not depend on $v_1$.}
Then, the line element in region II is written as
\be
	\dd s_\mathrm{II}^2
	=
	v_1^2 e^{2\zeta}
	\left(
		\frac{h(z)}{z} \dd \zeta^2
		-
		\frac{2}{z^2} \dd \zeta \dd z
		+
		\frac{1}{z^2} \dd \Omega_{d-2}^2
	\right)\ ,
\label{eq:zetar}
\ee
where
\be
	h(z)
	:=
	2 \mu z^{d-2} - z + 2\ .
\label{eq:hz}
\ee
Furthermore, if we introduce
\be
	\eta
	:=
	\zeta - 2z_\ast\ ,
\;\;\;\;\;\;
	z_\ast
	=
	\int^z
	\frac{ \dd z }{ z h(z) }\ ,
\label{eq:z(eta,zeta)}
\ee
the line element (\ref{eq:zetar}) is written in double null form,
\be
	\dd s_\mathrm{II}^2
	=
	v_1^2 e^{2\zeta}
	\left(
		\frac{h}{z} \dd \zeta \dd \eta
		+
		\frac{1}{z^2} \dd \Omega_{d-2}^2
	\right)\ .
\ee
Here, $z$ is understood as an implicit function of $\zeta$ and $\eta$ via Eq.~(\ref{eq:z(eta,zeta)}).
By a simple argument (see, \eg, \cite{JoshiBook}), one can show that the singularity appearing at $(v,r)=(0,0)$ is naked when the algebraic equation $h(z)=0$ [see Fig.~\ref{fg:hz} (a)] has two distinct positive roots, which we denote by $z_-$ and $z_+$ ($0 < z_- < z_+$).
For example, in $d=4$ the roots can be obtained explicitly as $z_{\pm}=( 1\pm \sqrt{1-16\mu})/4\mu$ for $0 < \mu < 1/16$. In general dimensions, the two positive roots exist if the accretion parameter $\mu$ is in the range of
\be
	0
	<
	\mu
	<
	\mu_c
	:=
	\frac{ (d-3)^{d-3} }{ [ 2(d-2) ]^{d-2} }\ ,
\label{eq:muc}
\ee
and in this case $z_\pm$ is restricted as
\be
	2 < z_- < z_c < z_+\ ,
\;\;\;\;\;\;
	z_c
	:=
	\frac{2(d-2)}{d-3}\ .
\label{eq:zc}
\ee
It turns out that $z=z_-$ and $z=z_+$ corresponds to the Cauchy horizon and event horizon, respectively. A conformal diagram of the Vaidya solution with $\mu$ satisfying condition (\ref{eq:muc}) is drawn in Fig.~\ref{fg:hz} (b).

\subsection{Connecting Cauchy horizon to outer event horizon}
\label{sec:matching}

In order to obtain a marginally naked-singular spacetime, the Cauchy horizon in region II, $ z=z_- $, is connected to the event horizon in the region III, $r = r_H := (2M)^{1/(d-3)} $, at the end of accretion of the null fluid, $v=v_0$. That is, we impose the following relation among three parameters ($\mu, v_0, M$),
\be
	z_-
	=
	\frac{ v_0 }{ ( 2M )^{1/(d-3)} }\ .
\label{eq:matching}
\ee


In this case, one can show that mass parameter $M$ is greater than the final quasi-local mass of the Vaidya solution, $\mu v_0^{d-3}$, as follow. From the matching condition (\ref{eq:matching}) and the conditions of $z_- < z_c$ and $\mu < \mu_c$, given in Eqs.~(\ref{eq:muc}) and (\ref{eq:zc}) respectively, we have
\be
	z_-^{d-3}
	=
	\frac{1}{2\mu} \cdot \frac{ \mu v_0^{d-3} }{ M } 
	<
	z_c^{d-3}
\;\;\;\;\;\;
\Longrightarrow
\;\;\;\;\;\;
	\frac{ \mu v_0^{d-3} }{ M }
	<
	2\mu z_c^{d-3}
	<
	2\mu_c z_c^{d-3}\ .
\label{eq:inequality-1}
\ee
With the help of explicit expressions of $\mu_c$ and $z_c$, Eq.~(\ref{eq:inequality-1}) leads to
\be
	M - \mu v_0^{d-3}
	>
	( d-3 ) \mu v_0^{d-3}
	>
	0\ ,
\;\;\;\;\;\;
	( 0 < {}^\forall \mu < \mu_c)\ ,
\ee
which is the desired inequality. Thus, the matching condition (\ref{eq:matching}) and the marginally naked singularity is realized, \eg, by infalling a shell of mass $M-\mu v_0^{d-3}$ toward the center at the moment of $v=v_0$. 

\begin{figure}[t]
	\begin{center}
			\includegraphics[width=6cm]{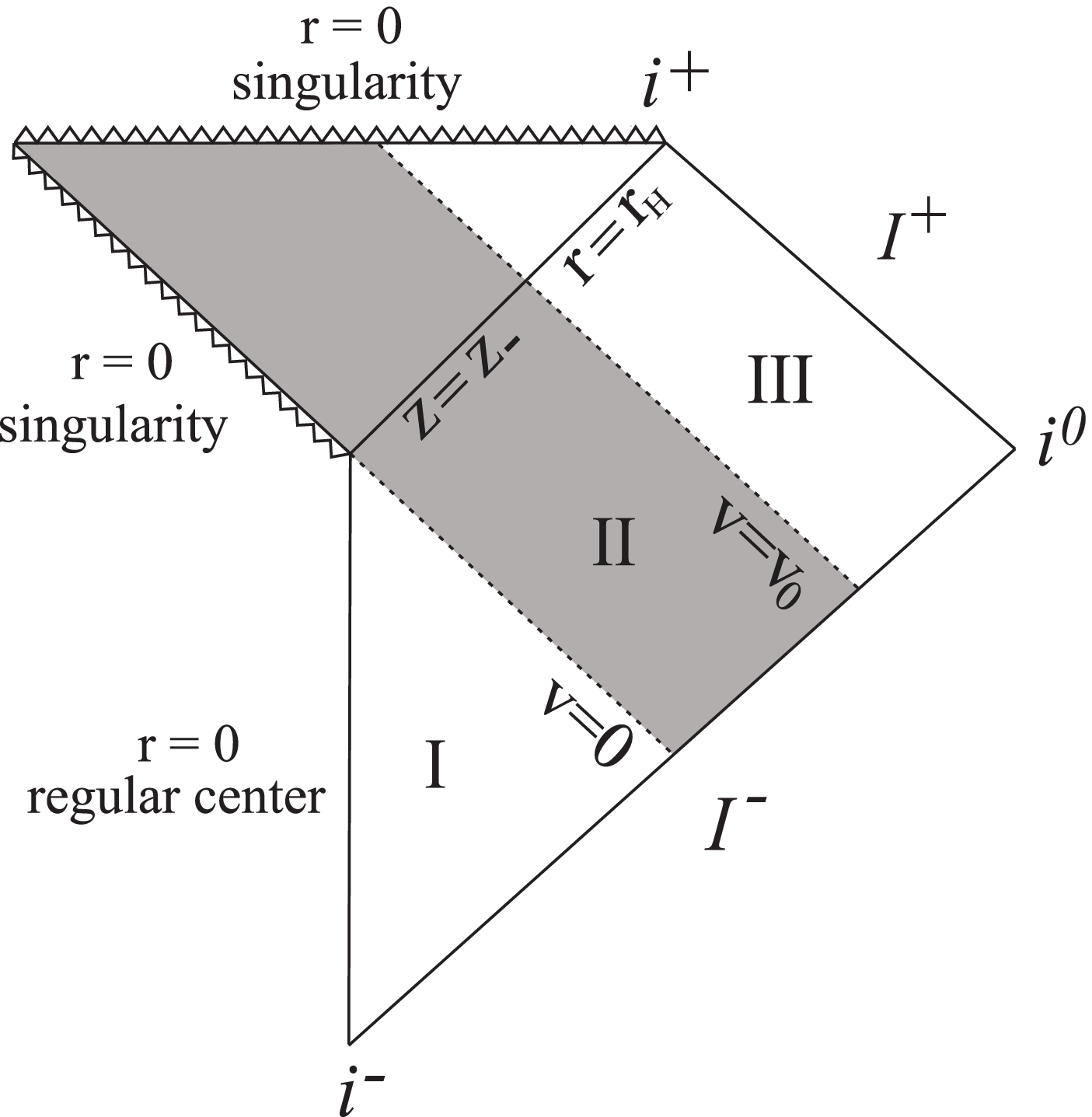} 
	\caption{\textsf{A conformal diagram of the spacetime considered in this paper. The Cauchy horizon ($z=z_-$) in the Vaidya region II is connected to the event horizon ($r=r_H$) in the Schwarzschild region III at $v=v_0$ by imposing condition (\ref{eq:matching}).}}
	\label{fg:conformal}
	\end{center}
\end{figure}

\section{Particle creation}
\label{sec:KG}

\subsection{Near-horizon behavior of massless scalar field}
\label{sec:nh}

The Klein-Gordon equation for a massless scalar field is given by
\be
	\dalm \phi
	=
	\frac{1}{\sqrt{-g}}\pd_a
	\left(
		\sqrt{-g} g^{ab} \pd_b \phi
	\right) = 0\ .
\ee
In the rest of this subsection we consider this equation in region II.
Separation of variables is possible in the ($\zeta,z$)-coordinates (\ref{eq:zetar}) by writing
\be
	\phi
	=
	e^{ \sigma \zeta/2 } Q(z) Y_{d-2}\ ,
\ee
where $\sigma$ is a constant of separation. $Y_{d-2}$ is a harmonic function on the unit ($d-2$)-sphere that satisfies
\be
	\big[
		\Delta_{d-2}
		+
		\ell ( \ell + d-3 )
	\big]
	Y_{d-2} = 0\ ,
\;\;\;\;\;\;
	\ell = 0,1,2,\ldots\ ,
\ee
where $\Delta_{d-2}$ is the Laplacian on the sphere.
The equation for $Q(z)$ takes form of an eigenvalue problem for $\sigma$,
\be
	zh \pd_z^2 Q
	+
	\left[
		2\mu z^{d-2} + (d-4)z + \sigma - (d-4)
	\right] \pd_z Q
	+
	\left[
		\ell( \ell + d-3 ) - \frac{ d-2 }{ 2z } \sigma
	\right] Q
	=
	0\ ,
\label{eq:Qeq}
\ee
where we have partially used the explicit expression of $h(z)$, Eq.~(\ref{eq:hz}).

While it seems difficult to obtain an analytic solution to Eq.~(\ref{eq:Qeq}) globally, it is enough for our aim to solve it near the event horizon, $z=z_-$, and the boundary between the regions I and II, $z=0$. To consider Eq.~(\ref{eq:Qeq}) near $z=z_-$, it is useful to introduce $\hat{h}(z)$ by
\be
	h(z) = (z-z_-) \hat{h}(z)\ ,
\;\;\;\;\;\;
	\hat{h}(z_-) \neq 0\ ,
\ee
where the non-vanishing of $\hat{h}(z_-)$ corresponds to that $h$ has a simple zero at $z_-$. Then, one can show that Eq.~(\ref{eq:Qeq}) has power solutions near $z = z_-$,
\be
	Q \simeq (z-z_-)^{s_1^\pm},
\;\;\;\;\;\;
	s_1^+
	=
	0\ ,
\;\;\;\;\;\;
	s_1^-
	=
	\frac{ \sigma + (d-2) }{ 2(d-2) - (d-3)z_- }\ ,
\label{eq:power1}
\ee
where we have used l'H\^opital's
rule to eliminate $\hat{h}(z_-)$ in the power,
\be
	\hat{h}(z_-)
	=
	\lim_{z \to z_-}
	\frac{ h(z) }{ z-z_- }
	=
	\lim_{z \to z_-} h'(z)
	=
	\frac{ (d-3)z_- - 2(d-2) }{ z_- }\ .
\ee
On the other hand, Eq.~(\ref{eq:Qeq}) has power solutions near the I-II boundary, $z = 0$, given by
\be
	Q \simeq z^{s_2^\pm},
\;\;\;\;\;\;
	s_2^+ = \frac{d-2}{2}\ ,
\;\;\;\;\;\;
	s_2^- = - \frac{\sigma}{2}\ .
\label{eq:power2}
\ee

\subsection{Quasi-thermal spectrum}
\label{sec:spectrum}

Now, we quantize the massless scalar field of which classical behaviors in region II were investigated above, and calculate the spectrum of particle creation largely according to Hiscock \et~\cite{Hiscock:1982pa}.

A mode function which defines the positive energy particles ($\pd_t p_\omega = -i\omega p_\omega$, $\omega > 0$) at the future null infinity $I^+$ is written as
\be
	p_\omega
	=
	\frac{ N }{ \sqrt{\omega} }
	\frac{ P_\omega (r) }{ r^{(d-2)/2} }
	e^{ -i \omega u }\ ,
\label{eq:p}
\ee
where $N$ is a normalization constant, and $u$ is the retarded time in the region III:
\be
	u
	:=
	v - 2r_\ast\ ,
\;\;\;\;\;\;
	r_\ast
	=
	\int^r
	\frac{ \dd r }{ 1 - (r_H/r)^{d-3} }\ .
\ee
Hereafter, we omit the dependence of mode functions on the spherical coordinates, \ie, $Y_{d-2}$, since it plays no important role in our analysis. $P_\omega(r)$ is a function that approaches the unity at the asymptotic region, $\lim_{r\to \infty} P_\omega(r)= 1$.
One the other hand, the mode function which defines the particle at the past null infinity $I^-$ is written as
\be
	f_{\omega^\prime}
	=
	\frac{ N^\prime }{ \sqrt{\omega^\prime} }
	\frac{ F_{\omega^\prime} (r) }{ r^{(d-2)/2} }
	e^{ -i \omega^\prime v },
\label{eq:f}
\ee
where $N^\prime$ is a normalization constant and $\lim_{r\to \infty} F_{\omega^\prime} (r) = 1 $ is assumed.

Since both $p_\omega$ and $f_{\omega^\prime}$ are complete sets, one of them can be expressed as a sum over the frequencies of the other. Namely,
\be
	p_\omega
	=
	\int_0^\infty
	(
		\alpha_{\omega \omega^\prime} f_{\omega^\prime}
		+
		\beta_{\omega \omega^\prime} \bar{f}_{\omega^\prime}
	) \dd \omega^\prime\ ,
\label{eq:pf-rel}
\ee
where $\alpha_{\omega \omega^\prime}$ and $\beta_{\omega \omega^\prime}$ are constants (the Bogoliubov coefficients) and the bar represents the complex conjugate.
If we assume that the quantum state is in the vacuum $|0\ran$ defined by $f_{\omega^\prime}$ at $I^-$ and observe the particles defined by $p_\omega$ at $I^+$, we have the number density given by
\be
	\lan 0 | N_\omega | 0 \ran
	=
	\int_0^\infty
	| \beta_{\omega \omega^\prime} |^2 \dd  \omega^\prime\ ,
\label{eq:N}
\ee
where $N_\omega$ is the particle number operator corresponding to $p_\omega$.
Thus, what to do in order to estimate the number density spectrum (\ref{eq:N}) is to propagate $p_\omega$ to $I^-$ and calculate the Bogoliubov coefficient, $\beta_{\omega \omega^\prime}$.

By tracing $p_\omega$ along the event horizon to the II-III boundary, $v=v_0$, we find
\be
	p_\omega
	\simeq
	\frac{ K_0 }{ \sqrt{ \omega } } e^{2i\omega r_\ast}\ ,
\label{eq:p0}
\ee
where $K_0$ is the product of the normalization constant, a scattering amplitude for the static geometry in region III, and the asymptotic limit of the factor $P_\omega/r^{(d-2)/2}$ at $r \to r_H$.
Writing $r_\ast$ in terms of $r$, Eq.~(\ref{eq:p0}) is written at the leading order,
\be
	p_\omega
	\simeq
	\frac{ K_1 }{ \sqrt{ \omega } }
	\left(
		\frac{ r }{ r_H } - 1
	\right)^{i\omega/\kappa}\ ,
\label{eq:p1}
\ee
where unimportant factors are absorbed in $K_1$, and $\kappa$ is the surface gravity,
\be
	\kappa
	:=
	\frac12 \pd_r
	\left.
	\left(
		1 - \left( \frac{r_H}{r} \right)^{d-3}
	\right)
	\right|_{r=r_H}
	=
	\frac{ d-3 }{ 2r_H }\ .
\ee
Furthermore, writing $r$ in terms of $z$, Eq.~(\ref{eq:p1}) becomes
\be
	p_\omega
	\simeq
	\frac{ K_2 }{ \sqrt{ \omega } }
	\left(
		z_- - z
	\right)^{ i\omega/\kappa }\ ,
\label{eq:p2}
\ee
where unimportant factors are absorbed in $K_2$.
Now, we would like to relate $p_\omega$ given by Eq.~(\ref{eq:p2}) with the the general solution given by Eq.~(\ref{eq:power1}). Clearly, mode function (\ref{eq:p2}) corresponds to the solution with the power $s_1^-$ in Eq.~(\ref{eq:power1}). Thus, we have a relation between $\omega$ and $\sigma$
\be
	\sigma
	=
	- (d-2)
	+ i \chi \omega\ ,
\;\;\;\;\;\;
	\chi
	:=
	\frac{ 2(d-2) - (d-3)z_- }{ 2\kappa }\ .
\label{eq:sigma}
\ee

From Eq.~(\ref{eq:power2}), the general solution near the I-II boundary, $v=0$, can be written as
\be
	\phi
	\simeq
	A e^{\sigma\zeta/2} z^{-\sigma/2}
	+
	B e^{\sigma\zeta/2} z^{(d-2)/2}
	=
	A r^{\sigma/2}
	+
	B \frac{ v^{\sigma/2 + (d-2)/2} }{ r^{(d-2)/2} }\ ,
\label{eq:general-sol}
\ee
where $A$ and $B$ are constants, and we have set $v_1$ to unity for notational simplicity.
Thus, if we trace the wave (\ref{eq:p2}) toward the neighborhood of $v=0$, and put it in the form of general solution (\ref{eq:general-sol}), it should take form of
\be
	p_\omega
	\simeq
	A_\omega \frac{  K_2}{ \sqrt{\omega} }
	\frac{ r^{ i\omega\chi } }{ r^{(d-2)/2} }
	+
	B_\omega \frac{ K_2}{ \sqrt{\omega} }
	\frac{ v^{ i\omega\chi } }{ r^{(d-2)/2} }\ ,
\label{eq:p3}
\ee
where we have expressed $\sigma$ in terms of $\chi$ through Eq.~(\ref{eq:sigma}).
The first term in Eq.~(\ref{eq:p3}) corresponds to the ingoing wave at $v=0$, while the second term corresponds to the wave scattered to $I^-$ near the singularity. Thus, $A_\omega$ and $B_\omega$ are the transmission and scattering coefficients, respectively, for the dynamical region II. They are related each other by the probability conservation $ | A_\omega |^2 - | B_\omega |^2 = 1 $.

Tracing the wave (\ref{eq:p3}) to $I^-$ via $v=0$ and then via the center $r=0$, mode function $p_\omega$ at $|v| \simeq 0$ is found to be
\be
	p_\omega
	\simeq
	\left\{
	\begin{array}{ll}
	{\displaystyle
		A_\omega^\prime \frac{  K_2 }{ \sqrt{\omega} }
		\frac{ e^{ i\omega\chi \ln |v| } }{ r^{(d-2)/2} }
	}\ ,
		& v < 0 \\
	{\displaystyle
		B_\omega  \frac{ K_2 }{ \sqrt{\omega} }
		\frac{ e^{ i\omega\chi \ln v } }{ r^{(d-2)/2} }
	}\ ,
		& v > 0 \\

	\end{array}
	\ \right.\ ,
\label{eq:p4}
\ee
where a phase shift of the transmitted wave is absorbed in $A_\omega^\prime$.

Now, we are ready to calculate the Bogoliubov coefficients with using Eq.~(\ref{eq:p4}). The coefficients are given by reversing Eq.~(\ref{eq:pf-rel}),
\be
	\alpha_{\omega\omega^\prime}
	=
	\sqrt{\omega^\prime} r^{(d-2)/2}
	\int_{-\infty}^\infty e^{i\omega^\prime v} p_\omega \dd v\ ,
\;\;\;\;\;\;
	\beta_{\omega\omega^\prime}
	=
	\sqrt{\omega^\prime} r^{(d-2)/2}
	\int_{-\infty}^\infty e^{-i\omega^\prime v} p_\omega \dd v\ .
\label{eq:coeff1}
\ee
Substituting $p_\omega$ in Eq.~(\ref{eq:p4}) into Eq.~(\ref{eq:coeff1}), we have
\be
	\alpha_{\omega\omega^\prime}
	=
	K_2
	\sqrt{ \frac{\omega^\prime}{ \omega } }
	\left(
		A^\prime_\omega J_- + B_\omega J_+
	\right)\ ,
\;\;\;\;\;\;
	\beta_{\omega\omega^\prime}
	=
	K_2
	\sqrt{ \frac{\omega^\prime}{ \omega } }
	\left(
		A^\prime_\omega J_+ + B_\omega J_-
	\right)\ ,
\ee
where
\be
	J_\pm
	:=
	\int_0^\infty
	\exp
	\left(
		\pm i\omega^\prime v + i \omega \chi \ln v 
	\right) \dd v\ .
\ee
This integration is nothing but the one evaluated by Hawking~\cite{Hawking:1974sw}. Thus, the resultant expressions of $(\alpha_{\omega\omega^\prime},\beta_{\omega\omega^\prime})$ are given as
\be
	\alpha_{\omega\omega^\prime}
	=
	\gamma_{\omega\omega^\prime}
	\left(
		- A^\prime_\omega e^{\pi \omega \chi/2}
		+ B_\omega e^{-\pi \omega \chi/2}
	\right)\ ,
\;\;\;\;\;\;
	\beta_{\omega\omega^\prime}
	=
	\gamma_{\omega\omega^\prime}
	\left(
		  A^\prime_\omega e^{-\pi \omega \chi/2}
		- B_\omega e^{\pi \omega \chi/2}
	\right)\ ,
\label{eq:coeff2}
\ee
where
\be
	\gamma_{\omega\omega^\prime}
	:=
	i K_2 \sqrt{ \frac{\omega^\prime}{ \omega } } 
	\Gamma( 1+i\omega\chi )
	\exp
	\left[
		-( 1+i\omega\chi ) \ln \omega^\prime
	\right]\ .
\ee
From Eq.~(\ref{eq:coeff2}), we obtain a relation
\be
	| \alpha_{\omega\omega^\prime} |^2
	=
	\frac{ | C_\omega -  e^{\pi\omega\chi} |^2 }
		 { | 1 - C_\omega e^{\pi\omega\chi} |^2 }
	| \beta_{\omega\omega^\prime} |^2\ ,
\;\;\;\;\;\;
	C_\omega
	:=
	\frac{ B_\omega }{ A^\prime_{\omega} }\ .
\label{eq:alpbeta-rel}
\ee
Conservation of probability may be expressed as
\be
	\int_0^\infty
	\left(
		| \alpha_{\omega\omega^\prime} |^2
		-
		| \beta_{\omega\omega^\prime} |^2
	\right) \dd \omega^\prime
	=
	\Gamma_\omega\ ,
\label{eq:prop-cons}
\ee
where $\Gamma_\omega$ is the transmission coefficient for the wave propagating backward from $I^+$ through the static geometry in region III and the low-curvature portion of region II. Now, we obtain the spectrum (\ref{eq:N}) with using Eqs.~(\ref{eq:alpbeta-rel}) and (\ref{eq:prop-cons}),
\be
	\lan 0 | N_\omega | 0 \ran
	=
	\frac{ \Gamma_\omega }{ e^{\omega/T_{\rm eff}} - 1 }\ ,
\;\;\;\;\;\;
	T_{\rm eff}
	:=
	\frac{ \omega }
		 { \ln \left[
					1
					+
					( e^{2\pi\omega\chi}-1 )
					/
					| 1 - C_\omega e^{\pi\omega\chi} |^2
			   \right] }\ ,
\label{eq:spectrum}
\ee
where we have introduced an frequency-dependent effective temperature to put the spectrum in the quasi-thermal form.

\section{Modified temperature and lifetime}
\label{sec:temperature}

\begin{table}[b!]
\begin{center}
\caption{Hawking temperature $T_H$ and its corresponding evaporation time $\tau_H$ when $M_P c^2 = 1 \; {\rm TeV}$ and $M_{BH} c^2 = 5 \; {\rm TeV}$. For reference, Schwarzschild radius $r_H$ is given in the unit of fermi (1 fm := $10^{-15}$ m).}
\label{tbl:temperature}
\vspace{1mm}
\setlength{\tabcolsep}{15pt}
\begin{tabular}{cccccccc}
\hline\hline
$d$ & $5$ & $6$ & $7$ & $8$ & $9$ & $10$ & $11$
\\
\hline
$k_B T_H$ (GeV) &  77.2 & 179 & 282 & 380  & 470  & 553 & 629 \\
$\tau_H$ ($10^{-26}$ sec) & 220 & 24.5 & 5.31 & 1.44 & 0.418 & 0.123 & 0.0357  \\
$r_H$ ($10^{-4}$ fm) &  4.07 & 2.64 & 2.23 & 2.07  & 2.01  & 1.99 & 2.00 
\\ \hline\hline
\end{tabular}
\end{center}
\end{table}

\begin{figure}[t!]
	\begin{center}
		\setlength{\tabcolsep}{25pt }
		\begin{tabular}{ cc }
			\includegraphics[width=7cm]{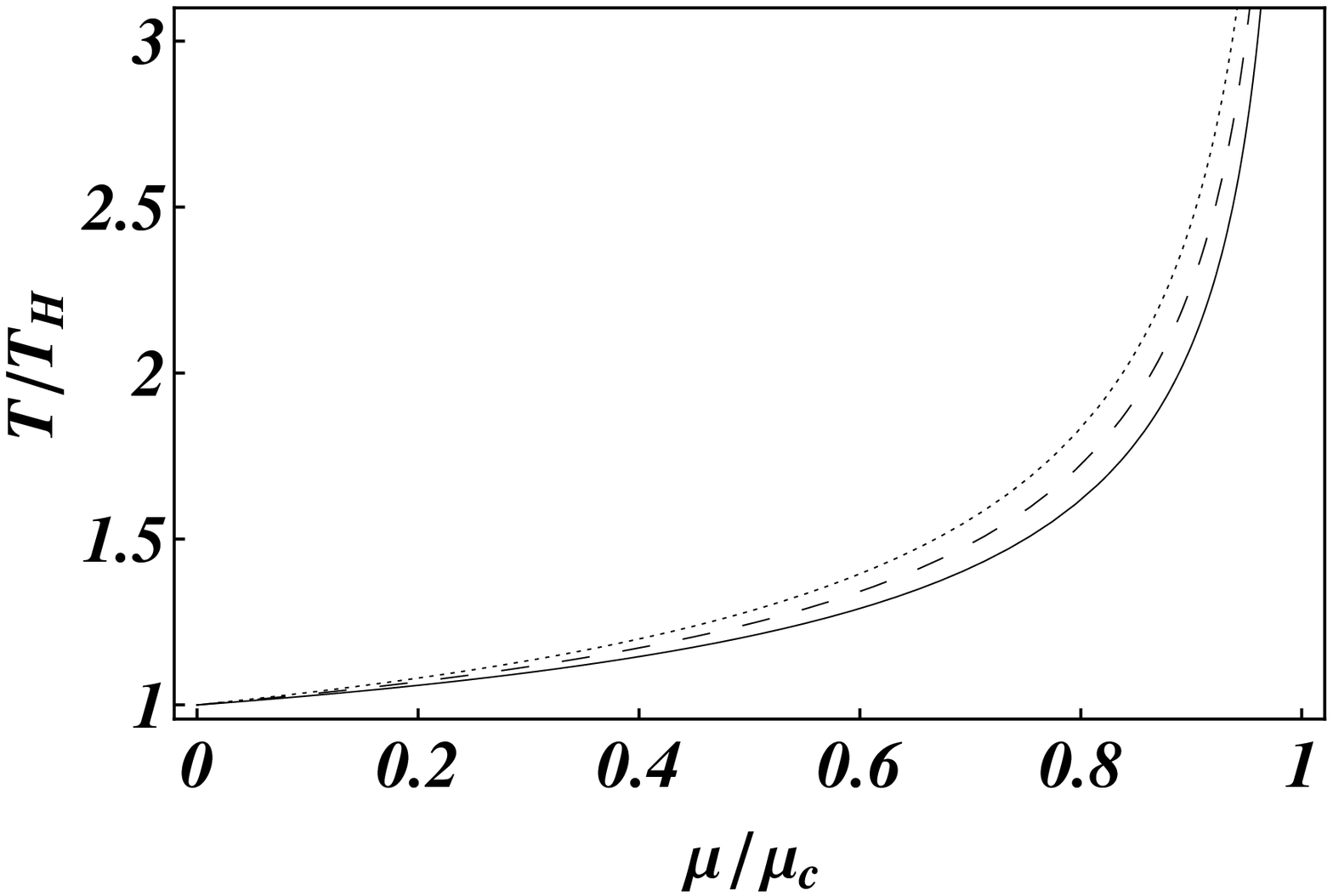} &
			\includegraphics[width=7cm]{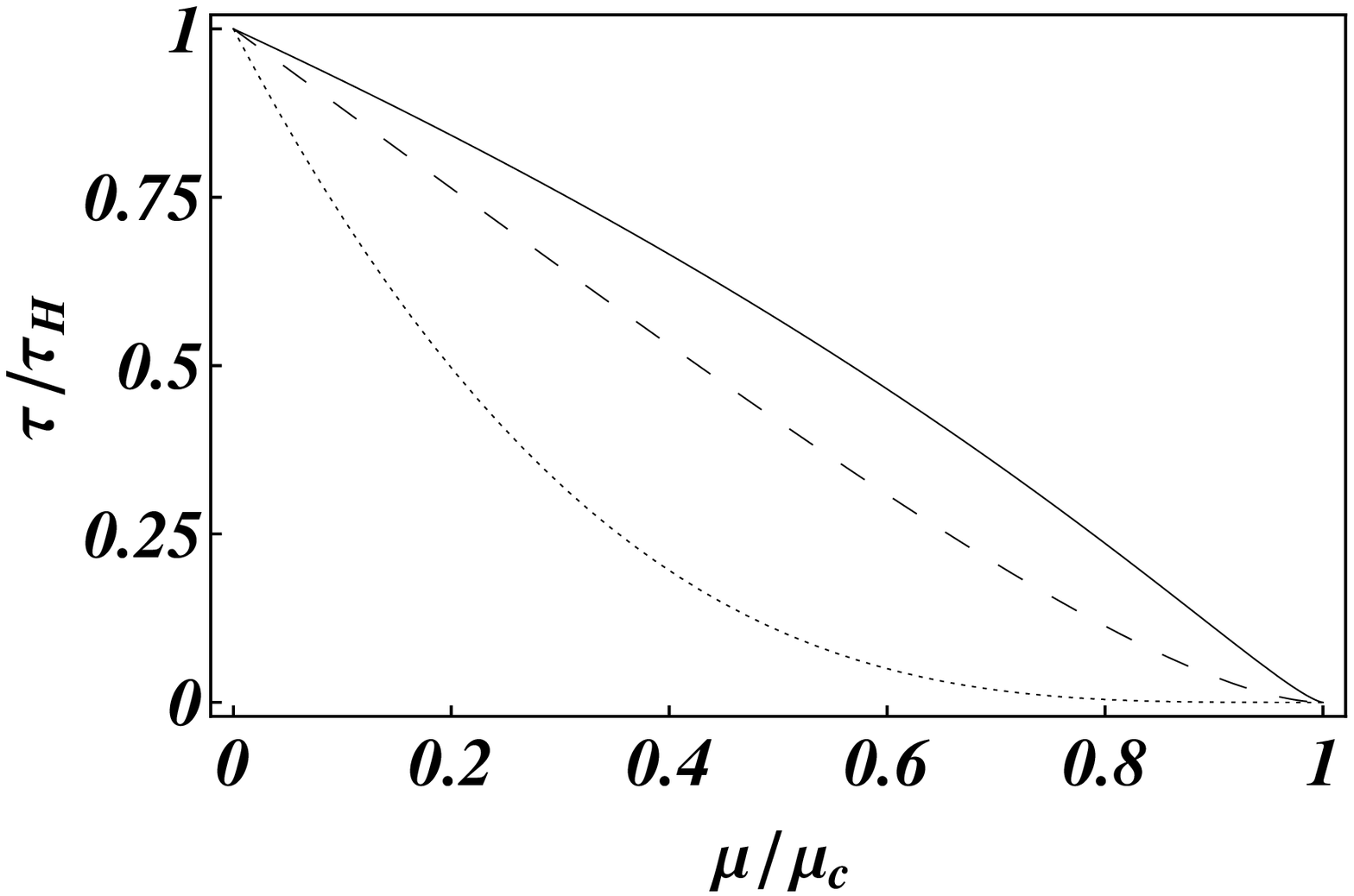} \\
			(a) & (b)
		\end{tabular}
	\caption{\textsf{(a) Accretion parameter $\mu$ \vs~temperature $T$ normalized by the Hawking temperature $T_H$ [see Eq.~(\ref{eq:T0})] for $d=4$ (solid), $5$ (dashed), and $10$ (dotted). The temperature diverges in the limit of $\mu \to \mu_c$. (b) Accretion parameter \vs~evaporation time normalized by that determined by the Hawking temperature [see Eq.~(\ref{eq:tau2})].}}
	\label{fg:temp}
	\end{center}
\end{figure}

In order to see the physical meaning of the effective temperature (\ref{eq:spectrum}), let us neglect the scattering near the singularity (\ie, take the limit of $C_\omega \to 0$) and define a temperature in this limit,
\be
	T
	:=
	\lim_{C_{\omega} \to 0} T_{\rm eff}
	=
	\frac{1}{2\pi \chi}
	=
	\frac{ 2 }{ (d-3) ( z_c - z_- ) } T_H\ ,
\;\;\;\;\;\;
	T_H
	=
	\frac{ d-3 }{ 4\pi r_H }\ ,
\label{eq:T0}
\ee
where $T_H = \kappa/2\pi$ is the Hawking temperature.
Reviving natural constants $(c,\hbar,k_B,G_d)$, $T_H$ is written in term of $M_{BH}$,
\be
	k_B T_H
	=
	M_{P}c^2
	\left(
		\frac{ M_{P} }{ M_{BH} }
	\right)^{1/(d-3)}
	\left(
		\frac{ (d-2)(d-3)^{d-3} }{ 2^{2d-3} \pi^{(d-3)/2} \Gamma[(d-1)/2] }
	\right)^{1/(d-3)}\ ,
\label{eq:TH}
\ee
where $ M_{P} := (\hbar^{d-3}/G_d c^{d-5})^{1/(d-2)} $ is the $d$-dimensional Planck mass.
Numerical values of Hawking temperature (\ref{eq:TH}) when $M_P c^2 = 1 \; {\rm TeV}$ and $ M_{BH} c^2 = 5 \; {\rm TeV}$ are given in Table~\ref{tbl:temperature}. Since $T$ is an increasing function of $z_-$, its minimum value is realized in the limit of $z_- \to 2$ or equivalently $\mu \to 0$. Namely, $ T > \lim_{z_- \to 2} T = T_H$.
Thus, the temperature is always higher than the Hawking temperature. One can also show that in the limit that the two homothetic horizons coalesce (realized by $z_\pm \to z_c$ or $\mu \to \mu_c$) the temperature diverges, $\lim_{z_- \to z_c} T = + \infty$. The $\mu$-dependence of $T$ is shown in Fig.~\ref{fg:temp} (a).

If we take into account the backreactions of particle creation, the modification of temperature affects the evolution of the spacetime, \eg, the lifetime or evaporation time.
The mass decreasing rate is proportional to the luminosity, which is the product of horizon area $\mathcal{A}_H$ and flux $\propto T^d$ (\ie, Stephen-Boltzmann law in $d$-dimensional spacetimes),
\be
	- \frac{ dM_{BH} }{ dt }
	\propto
	\mathcal{A}_H T^{d}
	\propto
	M_{BH}^{-2/(d-3)}\ ,
\label{eq:dmdt}
\ee
where we have used $\mathcal{A}_H \propto r_H^{d-2} \propto M_{BH}^{(d-2)/(d-3)}$ and $T \propto M_{BH}^{-1/(d-3)}$.
Integrating (\ref{eq:dmdt}), the evaporation time $\tau$ is estimated as
\be
	\tau
	\propto
	M_{BH}^{(d-1)/(d-3)}
	\propto
	T^{-(d-1)}\ .
\label{eq:tau1}
\ee
Taking into account that Eq.~(\ref{eq:T0}) holds in our case, we obtain
\be
	\tau
	=
	\left[
		\frac{ (d-3)( z_c-z_- ) }{ 2 }
	\right]^{d-1}
	\tau_H\ ,
\label{eq:tau2}
\ee
where $\tau_H$ is the evaporation time of a black hole of temperature $T_H$,
\be
&&
	\tau_H
	=
	\frac{ \hbar }{ M_P c^2 }
	\left( \frac{M_{BH}}{M_P}  \right)^{(d-1)/(d-3)}
	\left(
		\frac{ 2^{2d^2-7d+9} \Gamma[(d-1)/2]^{d-1} }
			 { (d-2)^2 }
	\right)^{1/(d-3)}
\nn
\\
&&
\hspace{7cm}
	\times
		\frac{ \pi^{(2d-1)/2} }
	{ (d-3)^d (d-1)^2 \Gamma[(d+2)/2] \zeta(d) } \ .
\nn
\\
\ee
Here, $\zeta$ is the Riemann zeta-function. The numerical values of $\tau_H$ are given in Table~\ref{tbl:temperature}.\footnote{Here, $\tau_H$ is the evaporation time of a black hole in a minimal scenario. That is, we assume that only gravitons, of which polarization degrees of freedom is $(d-3)d/2$ in $d$-dimensions, propagate in the bulk (see the first paper in Ref.~\cite{Argyres:1998qn}).} Thus, the lifetime is always shorter than that of an ordinary black hole of the same mass, and even can be arbitrarily short in the limit that the two homothetic horizons coalesce. The $\mu$-dependence of $\tau$ is shown in Fig.~\ref{fg:temp} (b).

\section{Conclusion}
\label{sec:conclusion}

Motivated by the possible production of naked singularities by high-energy particle collisions in the context of TeV-scale gravity, we have investigated the particle creation by naked singularities in higher dimensions. For simplicity, the background spacetime has been modeled by the $d$-dimensional spherically symmetric collapsing solution, given by Eqs.~(\ref{eq:metric}) and (\ref{eq:m(v)}). In addition, the Cauchy horizon in the Vaidya region is matched to the event horizon in the outer region by imposing a particular condition (\ref{eq:matching}). This matching enables us to calculate the Bogoliubov coefficients without the ambiguity of boundary conditions at the singularity. We have found that the spectrum of the created particles deviates from that of the usual Hawking radiation due to the scattering near the singularity. However, the spectrum can be recast in the quasi-thermal form with a frequency-dependent temperature, Eq.~(\ref{eq:spectrum}). Neglecting the scattering near the singularity, the temperature is defined as Eq.~(\ref{eq:T0}) and compared with the Hawking temperature. The temperature is always higher than the Hawking temperature of a same-mass black hole, and therefore the evaporation time is always shorter than that of the same-mass black hole, in the whole range of parameters (\ie, spacetime dimensions $d$ and accretion parameter $\mu$). This is a concrete prediction about the particle creation during naked-singularity formation in higher dimensions, made by this paper for the first time (as far as the authors know).

One might wonder that matching condition (\ref{eq:matching}) choses a particular class of solutions out of more general solutions. While that condition is essential to avoid the ambiguity of boundary conditions at the singularity, there seems no physical reason why such a class of solutions with the marginally naked singularity are preferred. Therefore, a more general case including globally naked singularities should be analyzed. The present authors are working in this direction and will report it elsewhere soon~\cite{progress}.

Here, we discuss the relation between the analysis in this paper with that in~\cite{Harada:2000me}. Paper \cite{Harada:2000me} argued that (in the usual 4-dimensional gravity) if unknown quantum gravitational effects dominate near the birth of naked singularity to suppress the particle creation, the net energy radiated amounts only to a few Planck energy $M_{P(4)} \sim 10^{16} \; {\rm TeV}$. If this is the case, the particle creation by the naked singularity forming in the gravitational collapse of a {\it massive} object, \eg, an astrophysical star of mass $M_{\odot} \sim 10^{54} \; {\rm TeV}$, will be a negligible effect. However, we stress that in the TeV-scale gravity scenarios, even if the quantum gravitational cutoff works, the particle creation by the naked singularity can be a crucial effect. Namely, even if the net energy radiated by the particle creation amounts only to a few Planck energy $ \sim O({\rm TeV})$ due to the cutoff, the naked singularity produced by particle collisions (\eg, in nowadays collider experiments) itself would have the same order of mass at most. Therefore, the dynamics of background geometry is significantly affected by the particle creation\footnote{Incidentally, we should repeat here that the present model is quite simplified, and therefore the validity of the approximations used in this paper should be verified carefully when we discuss the actual process of visible-border production. For instance, the visible-border production is expected to be a highly dynamical process, of which typical timescale may be given by the light-crossing time through the collision region, and therefore the validity of the geometric-optics method and/or the QFT in curved background itself could be questionable. The authors thank an anonymous referee for focusing our attention to this point.}.

The production of naked singularities (or visible borders of spacetime) in the TeV-scale gravity is an intriguing possibility. However, we have not gotten any evidence of the production of a naked singularity nor a black hole in particle collisions so far. In recent years, the Numerical General Relativity is expanding to higher dimensions. In fact, several results on head-on collisions of black holes in higher dimensions have been reported ~\cite{Witek:2010xi}. (See \cite{Sperhake:2008ga} and \cite{Shibata:2008rq} for the high-energy collisions of 4-dimensional black holes with zero and non-zero impact parameters, respectively.) While the collision of higher-dimensional black holes would be different from that of Standard Model particles, which are confined to our 3-brane, it would be interesting to look for any signs or symptoms of naked-singularity/visible-border formation in numerical experiments.



\subsection*{Acknowledgments}

The authors thank T.~Harada, K.~Nakao, and anonymous referees for useful discussions and comments. Works of UM are partially supported by the Grant-in-Aid for Scientific Research Fund of the Ministry of Education, Culture, Sports, Science and Technology, Japan [Young Scientists (B) 22740176], and by Research Center for Measurement in Advanced Science in Rikkyo University.



\end{document}